\documentclass[aps, prl, superscriptaddress, twocolumn, amsfonts, amsmath, amssymb, floatfix]{revtex4-2}

\usepackage{graphicx}
\usepackage{float}
\usepackage{sidecap}
\usepackage{dcolumn}
\usepackage{bm}
\usepackage{epsfig}
\usepackage{braket}
\usepackage{color}
\usepackage{amsmath}
\usepackage[toc,page]{appendix}
\usepackage[colorlinks=true, letterpaper=true, pdfstartview=FitV, linkcolor=blue, citecolor=blue, urlcolor=blue]{hyperref}

\DeclareMathOperator{\sinc}{sinc}
\DeclareMathOperator{\rect}{rect}

\begin{document}

\title{Observation of Distinct Phase Transitions in a Nonlinear Optical Ising Machine}

\author{Santosh Kumar}
\email{skumar5@stevens.edu}
\author{Zhaotong Li}%
\author{Ting Bu}%
\author{Chunlei Qu}
\email{cqu5@stevens.edu}
 \author{Yuping Huang}%
\email{yhuang5@stevens.edu}

\affiliation{Department of Physics, Stevens Institute of Technology, Hoboken, NJ, 07030, USA}%
\affiliation{Center for Quantum Science and Engineering, Stevens Institute of Technology, Hoboken, NJ, 07030, USA}

\date{\today}

\begin{abstract}
Optical Ising machines promise to solve complex optimization problems with an optical hardware acceleration advantage.
Here we study the ground state properties of a nonlinear optical Ising machine realized by spatial light modulator, Fourier optics, and second harmonic generation in a nonlinear crystal. By tuning the ratio of the light intensities at the fundamental and second-harmonic frequencies, we experimentally observe two distinct ferromagnetic-to-paramagnetic phase transitions: a second-order phase transition where the magnetization changes to zero continuously and a first-order phase transition where the magnetization drops to zero abruptly as the effective temperature increases. Our experimental results are corroborated by a numerical simulation based on the Monte Carlo Metropolis-Hastings algorithm, and the physical mechanism for the distinct phase transitions can be understood with a mean-field theory. Our results showcase the great flexibility of the nonlinear optical Ising machine, which may find important potential applications in solving combinatorial optimization problems.
\end{abstract}

\maketitle

\textcolor{blue}{\textit{Introduction.---}}
Combinatorial optimization is ubiquitous and fundamental in many areas of science, engineering, finance, and social networks \cite{Mohseni2022}. Many optimization problems, such as the traveling salesman problem \cite{Laporte1992,Shaked:07}, the graph coloring problem \cite{parihar2017vertex}, the Boolean satisfiability problem \cite{Kirkpatrick1297}, spin glass dynamics \cite{bryngelson_spin_1987,Leonetti_21}, protein folding \cite{bryngelson_spin_1987}, etc., belong to the non-deterministic polynomial time (NP) hard or NP-complete class which can be formulated as finding the ground states of Ising spin models \cite{lucas_ising_2014,Kalinin2022}. Because of the computational complexity, it is usually challenging to find the exact solutions of general Ising models with traditional electronic computers \cite{Complexity_book,arora2009computational,Hamerly2019,Prabhu:20}. In the past years, many physical systems including the superconducting circuits \cite{leib_transmon_2016}, stochastic nanomagnets \cite{borders2019integer,Dutta2021}, trapped ions \cite{kim2010quantum}, complementary metal-oxide semiconductor devices  \cite{yamaoka201520k}, injection-locked laser networks  \cite{nixon_observing_2013}, polariton condensates \cite{ohadi_spin_2017,Kalinin2020}, etc., have been applied to realize an Ising simulator and to solve the optimization problems with heuristic search algorithms \cite{mahboob_electromechanical_2016,chou_analog_2019,Cai2020,Roques-Carmes2020,Parametric_Oscillators}. 

Among these physical implementations, optical Ising machines are particularly attractive because of their capability of parallelism, low energy consumption, and operation at the speed of light \cite{Marandi2014937,babaeian_single_2019,bohm_poor_2019,Roques-Carmes2020,Pierangeli2019,Goto2021,Honari-Latifpour2022}. In addition to the promising approach based on a network of degenerate optical parametric oscillators \cite{McMahon2016614,Inagaki2016603,Hamerly2019,Okawachi2020}, optical Ising machines based on spatial light modulators (SLM) and Fourier optics are also being actively pursued in the optical community \cite{Pierangeli2019,Kumar2020,pierangeli2020adiabatic,pierangeli2021scalable,PRLHuang2020,Huang2021,Sun:22}. By encoding the spins in the SLM-modulated binary phase of an incident beam and measuring the light intensity at the focal plane, a fully-connected large-scale optical Ising machine with configurable two-body spin-spin interactions can be realized. Furthermore, by including a second-harmonic (SH) light generation through nonlinear crystal and measuring the superposition of the pump light and SH light intensities, we recently have realized a more general Ising model with both two-body and four-body spin interactions \cite{Kumar2020}. 

A natural question that arises is whether such a nonlinear optical Ising machine can be used to solve optimization problems more efficiently. As the first step to address this important question, we experimentally and theoretically investigate the ground state magnetic phases of the nonlinear optical Ising machine for different four-body spin interaction coefficient and effective temperature. The main finding is that we can identify two distinct types of phase transitions with the order parameter - the magnetization - changes either continuously or abruptly to zero as the increase of temperature, corresponding to a second-order and a first-order phase transition respectively. We point out that similar Ising models with nearest-neighbor two-body and local four-body spin interactions have been theoretically explored in the 1970s \cite{Wu_PRB71,Lieb72,Oitmaa_1973}. To the best of our knowledge, our results represent the first experimental observation of two types of phase transitions in a configurable optical Ising model with fully connected two-body and four-body spin interactions.

\begin{figure}[t!]
\centering
\includegraphics[width=0.5\textwidth]{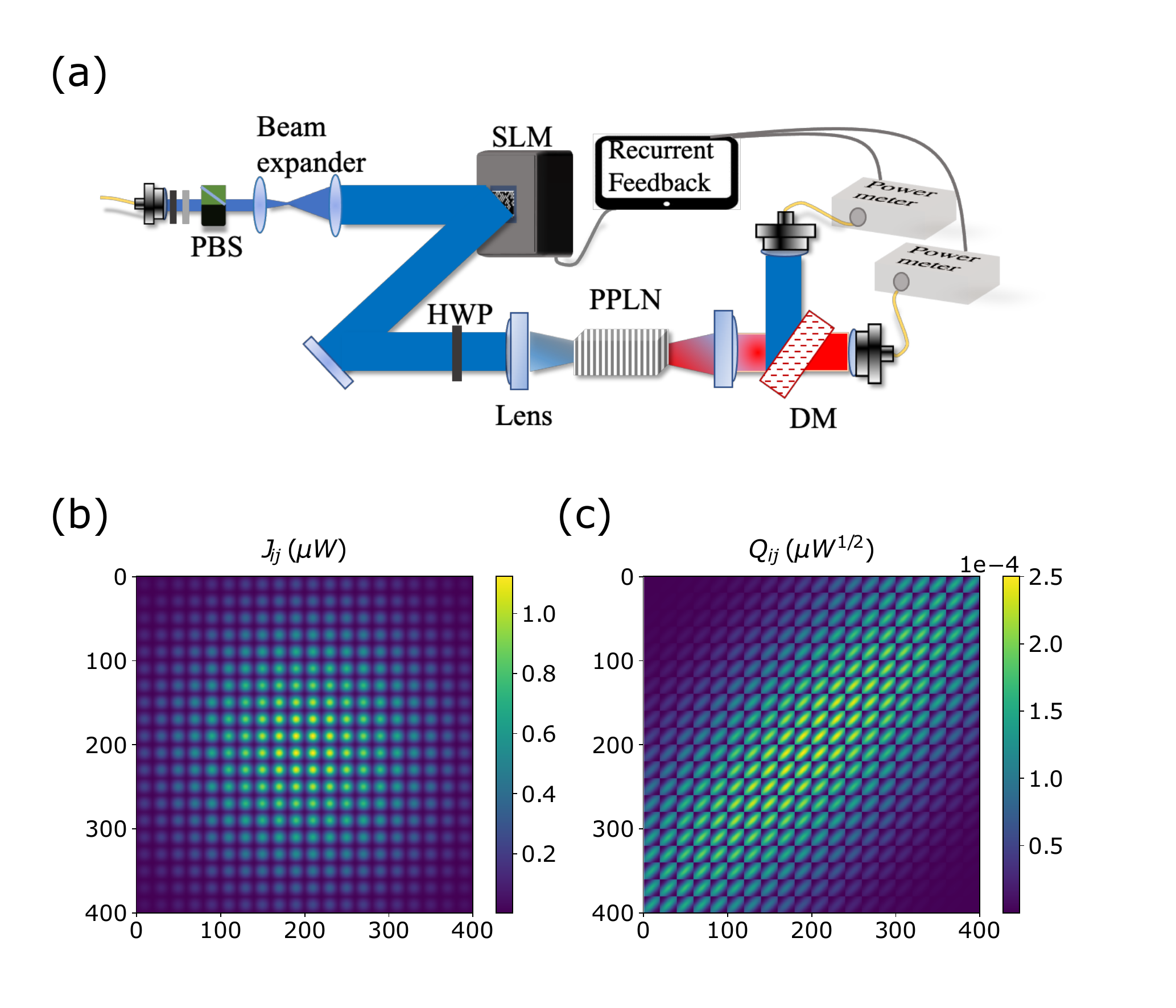}
\caption{(a) Sketch of the experimental setup for the nonlinear optical Ising machine. The linearly polarized pump light (blue) incidents on a SLM, and then it is focused into a nonlinear crystal (PPLN). The generated SH (red) and residual pump lights are coupled into separate single mode fibers and measured by the power meters. (b) The two-body and (c) the reduced four-body spin interaction matrices $J_{ij}$ and $Q_{ij}$ for a $20\times 20$ spin system.}
\label{fig:ExpSetup}
\end{figure}

\textcolor{blue}{\textit{Experimental setup.---}}
The schematic of the experimental setup for the nonlinear optical Ising machine is shown in Fig.~\ref{fig:ExpSetup}(a). We use a mode-locked laser of wavelength $\lambda \sim 1550.9$nm with an average power of 70mW as the pump light. The full width half maximum of the pump beam is $w_p=3.8$mm which is incident on the SLM (Santec SLM-100, 1440×1050 pixels, pixel length $a=10 \mu m$). The region of interest on the SLM is defined as a square lattice of $N=20\times 20$ giant spins with each spin consisting of $20\times 20$ pixels of the same phase which is modulated to be $0$ or $\pi$ to generate a random initial spin configuration. The unmodulated portion of the pump light is deflected by an optimized blazed grating. A lens with focal length $F=200$mm is used to focus the modulated beam into a temperature-stabilized periodically poled lithium niobate (PPLN) crystal with a poling period of $\Lambda$=19.36$\mu$m (5mol.\% MgO-doped PPLN, length 1cm from HC Photonics). It generates a SH light at $\lambda_h \sim 775.5$nm \cite{Bu:22}. After passing through another lens of the same focal length, the pump and SH lights are separated by a dichroic mirror (DM) and then coupled into the single mode fibers (SMF-28) using fiber collimators with aspheric lenses Thorlabs C220TMD-C and A375TM-B, respectively, and detected by the power meters (Thorlabs PM-100D with sensors S132C and S130C). The measurements are sent to the computer MATLAB interface, which completes the feedback loop by updating the SLM \cite{Kumar2020}. The spin flipping during each iteration is accepted or rejected according to a Boltzmann probability function $P=\exp(-\Delta U/T)$, where $\Delta U=E_{new}-E_{old}$ is the change in energy of a target Hamiltonian and $T$ is the effective temperature.

\textcolor{blue}{\textit{Theoretical Model.---}} Since the phase of the light incident on the SLM is modulated to be either $0$ or $\pi$, we can model the electric field of the incident light with the following discretized form 
\begin{eqnarray}
    E_p(\mathbf{x}) = \sum_i \xi_i \sigma_i \rect \left(\frac{\mathbf{x}-\mathbf{x}_i}{a}\right),
\end{eqnarray}
where $\mathbf{x}=(x, y)$ denotes the spatial coordinate on the SLM plane, $\mathbf{x}_i$ is the position of the $i^{th}$ pixel, $a$ is the pixel length, $\sigma_{i}=\pm 1$ is the phase of the pump light, and $\xi_{i}=E_0 e^{-(x_{i}^2+y_{i}^2)/w_p^2}$ corresponds to the amplitude of the Gaussian beam, $\text{rect}(\mathbf{x})=\rect(x)\rect(y)=1$ for $|x|<0.5$ and $|y|<0.5$ is the rectangular function. The wave reflected by the SLM passes through a lens of focal length $F$. On the focal plane, the wave is transformed to the Fourier domain according to
\begin{equation}
    \Tilde{E}_p(\mathbf{x}^\prime)=\frac{1}{F\lambda}\int E_p(\mathbf{x})e^{i\frac{2\pi}{F\lambda}\mathbf{x}\cdot\mathbf{x}^\prime}d\mathbf{x},
\end{equation}
where $\mathbf{x}^\prime=(x', y')$ denotes the spatial coordinate on the focal plane. A straightforward calculation gives
\begin{eqnarray}
    \tilde{E}_p(\mathbf{x}^\prime)
    &=&\frac{a^2}{F\lambda}\sum_{i}\xi_{i}\sigma_{i}e^{-i\frac{2\pi}{F\lambda}\mathbf{x}\cdot\mathbf{x}_i^\prime}\sinc\left(\frac{\pi a \mathbf{x}}{F\lambda}\right),
\end{eqnarray}
where $\sinc(\mathbf{x})=\sin(\pi x)\sin(\pi y)/(xy)$ denotes the two-dimensional sinc function.

The PPLN nonlinear crystal placed at the focal point of the lens generates SH light of frequency $\omega_h$. In the pump non-diffraction and non-depletion regime~\cite{boyd2020nonlinear}, the pump light in the nonlinear crystal does not change while the SH light can be obtained as $\tilde{E}_h(\mathbf{x}^\prime)=A_s\tilde{E}_p^2(\mathbf{x}^\prime)$
where $A_s=i\omega_h^2\chi^{(2)}L/(2\kappa_hc^2)$, $L$ is the length of the nonlinear crystal, $\chi^{(2)}$ is the second-order susceptibility, $c$ is the speed of light, $\kappa_h=2\pi n_h/\lambda_h$ is the wave number of the SH light, and $n_h$ is the index of refraction. After passing through two lenses of focal length $F$ and $f$, the pump light and SH light are subsequently coupled into the fibers and then the intensities are measured. The light intensity coupled into the fiber can be defined as
$P=\frac{1}{2}c\epsilon_0|\int E(\mathbf{u})E_f(\mathbf{u})d\mathbf{u}|^2$, where $\mathbf{u}=(u, v)$ and $E_f(\mathbf{u})=\sqrt{\frac{2}{\pi}}\frac{1}{w_f}\exp\left(-\frac{u^2+v^2}{w_f^2}\right)$ are the spatial coordinate and lowest optical mode (characterized by the width $w_f$ which is slightly different for the pump and SH lights) of the fiber, respectively. 

The Hamiltonian of our Ising spin model is defined as the superposition of the pump and SH light intensities,
\begin{eqnarray}
    H&=&-\sum_{i,j}J_{ij}\sigma_i\sigma_j-\gamma\sum_{i,j,s,r}K_{ijsr}\sigma_i\sigma_j\sigma_s\sigma_r, 
\end{eqnarray}
where we have multiplied $-1$ for pump light intensity (so that a ferromagnetic phase is favored at $T=0$) and $\gamma$ is a tunable parameter. It is clear that the two terms correspond to a two-body and a four-body spin interaction, with the following explicit expressions for the spin interactions 
\begin{eqnarray}
    J_{ij}&=&2\pi(w_f^p)^2\frac{a^4}{f^2\lambda^2}\xi_i\xi_j\xi_i^f\xi_j^f,\\
    K_{ijsr}&=&\frac{1}{2}\pi A_s^2(w_f^h)^2
    \frac{a^8}{F^2f^2\lambda^4}
    \xi_i\xi_j\xi_s\xi_r\xi_{ij}'^f\xi_{sr}'^f,
    \label{eq:K}
\end{eqnarray}
where $\xi_i^f=\exp\bigg[-\frac{\pi^2(w_f^p)^2}{f^2\lambda^2}(x_i^2+y_i^2)\bigg]$ and
$\xi_{ij}'^f=\exp\bigg[-\frac{\pi^2(\omega_f^h)^2}{f^2\lambda^2}[(x_i+x_j)^2+(y_i+y_j)^2]\bigg]$.

It is worth pointing out that in the pump non-diffraction and non-depletion regime, the two-body and four-body spin interactions are not full-rank matrices. They can be decomposed as $J_{ij}=P_iP_j$ with $P_i=\sqrt{2\pi}w_f^p a^2\xi_i\xi_i^f/f\lambda$ and $K_{ijsr}=Q_{ij}Q_{sr}$ with $Q_{ij}=\sqrt{\pi}A_sw_f^h a^4\xi_i\xi_j\xi_{ij}^f/\sqrt{2}Ff\lambda^2$. Thus, the two-body spin interaction $J_{ij}$ is a matrix of rank one and the reduced four-body interaction $Q_{ij}$ is a matrix of rank $N$ as the fiber mixes the light from different pixels (see Fig.~\ref{fig:ExpSetup}(b, c)). Such low-rank two-body and four-body spin interactions may impose some limitations to the SLM-based optical Ising machine for solving complex optimization problems. This can be circumvented, for example, by adding a scattering medium in front of the detector, as proposed in the recent work~\cite{pierangeli2021scalable,PRA_Gigan22}. Alternatively, the relative locations of the Fourier lenses can be changed to induce non-negligible diffraction inside the crystal, or the pump power can be increased to enter the pump-depletion regime. In either case, $J_{ij}$ and $K_{ijsr}$ will become full rank matrices and the above decomposition is not valid anymore. Furthermore, the ratio between the four-body and two-body spin interactions can also be tuned by changing other parameters, such as the pixel length $a$, the laser's beam width, etc., which may provide some additional flexibility for solving optimization problems with the nonlinear optical Ising machine.

\begin{figure}[t!]
\centering
\includegraphics[width=0.49\textwidth]{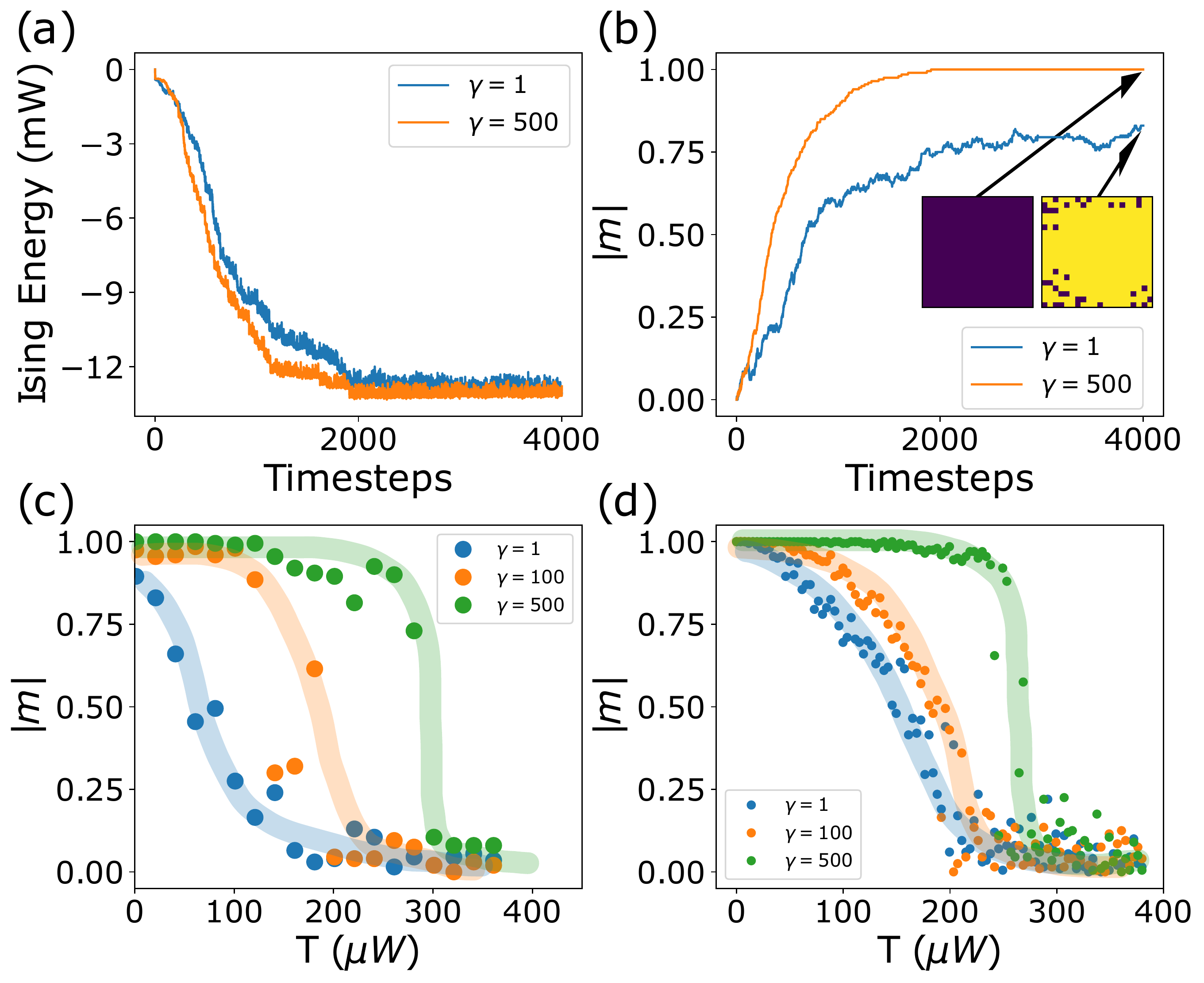}
\caption{(a) Experimentally measured Ising energy and (b) magnetization of the spin model as a function of the Monte-Carlo iterations for four-body coefficient $\gamma=1$ (blue) and $500$ (orange). (c) Experimentally measured magnetization as a function of temperature for $\gamma=0, 100, 500$. (d) The same results obtained from full Monte-Carlo numerical simulation. In (c, d), the magnetization drops to zero continuously at $\gamma=1, 100$ and abruptly at $\gamma=500$ with the thick lines the guides to the eyes.}
\label{fig:compare}
\end{figure}

\begin{figure}[t!]
\centering
\includegraphics[width=0.49\textwidth]{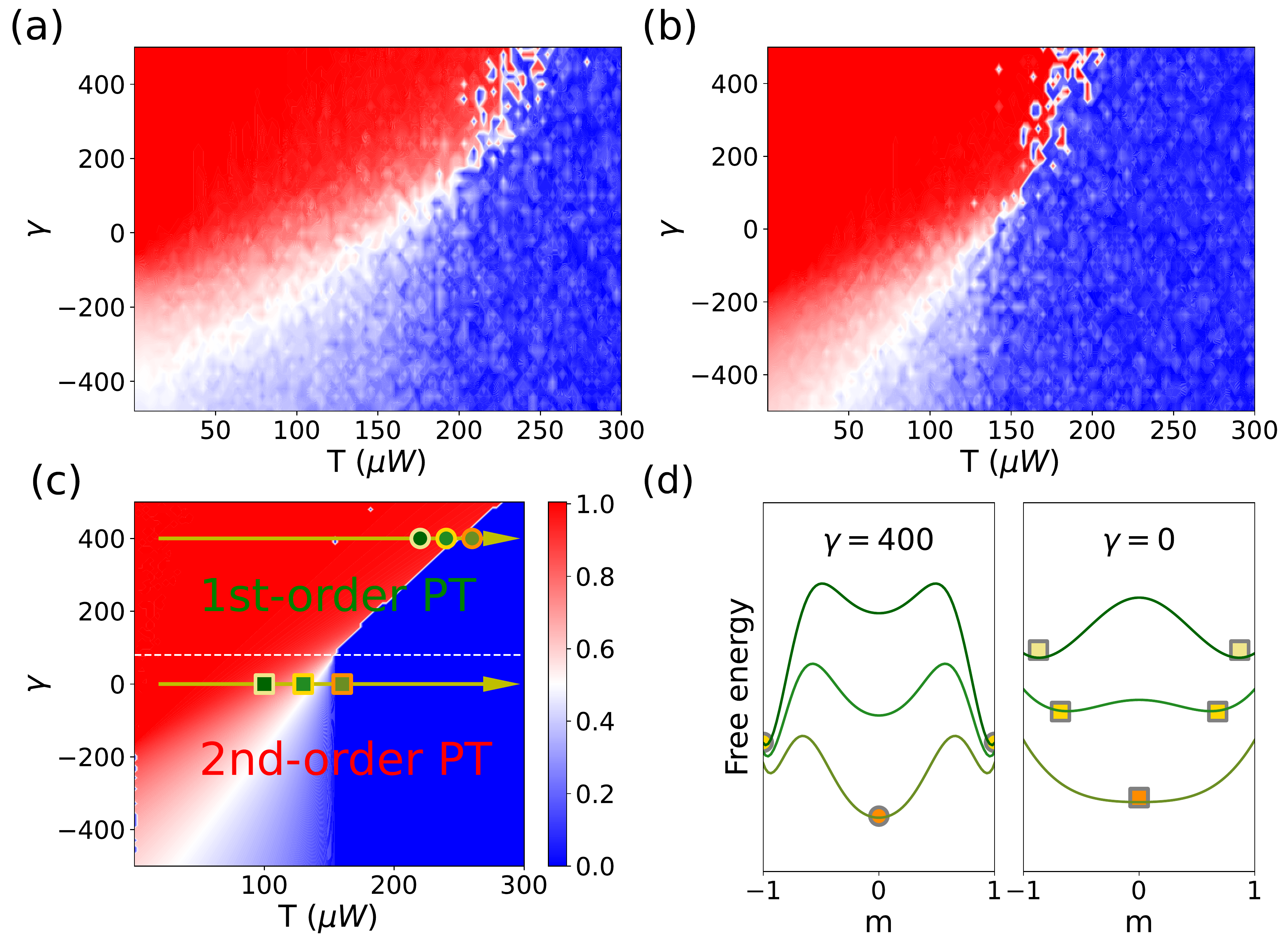}
\caption{Phase diagrams of the nonlinear optical Ising machine. (a, b) are the results obtained from the full Monte-Carlo numerical simulation with the actual nonuniform and approximated uniform spin-spin interactions, respectively, and (c) is the result obtained from the MF theory, where PT is phase transition. (d) The free energy as a function of magnetization at different temperature $T$ and four-body coefficient $\gamma$. The circle and square symbols correspond to the true ground state. All the three contour plots share the same color bar. In (b, c), we have used uniform two-body and four-body spin interactions with $J\approx 0.19\mu W$, $K\approx 3.6\times 10^{-9} \mu W$.}
\label{fig:MFT}
\end{figure}

\textcolor{blue}{\textit{Experimental results and numerical simulations.---}}Our main experimental observations are summarized in Fig.~\ref{fig:compare}(a, b, c). For a fixed effective temperature $T$, the Ising energy decreases to its lowest value after some Monte-Carlo iterations and the corresponding magnetization increases from zero (as we start with an initial random spin configuration) to a finite value that depends on the four-body coefficient. As shown in Fig.~\ref{fig:compare}(c), the magnetization changes to zero continuously as a function of temperature for $\gamma=0, 100$ while it drops to zero abruptly for $\gamma=500$, indicating two different types of phase transitions. These experimental observations can be reproduced with a full Monte-Carlo numerical simulation. Using the same Metropolis-Hasting algorithm \cite{liu_metropolis_2004}, we randomly choose one spin each time to flip in order to reduce the total energy. The spin flip is accepted with the Boltzmann probability, similar to that adopted in the experiment. This allows the system to evolve out of energy local minima and have more chance to reach the ground state. As shown in Fig.~\ref{fig:compare}(d), the numerical results are qualitatively the same as the experimental results. One of the mechanisms for the discrepancy is related to the optical loss, which is not captured in our theoretical model. By scanning the four-body coefficient $\gamma$ and effective temperature $T$, we numerically obtain the phase diagram as shown in Fig.~\ref{fig:MFT}(a). The phase diagram indicates that there is a second-order phase transition for small $\gamma$ and a first-order phase transition for large positive $\gamma$. We further assume a uniform two-body and four-body spin interaction constants and obtain a qualitatively similar phase diagram shown in Fig.~\ref{fig:MFT}(b).

\textcolor{blue}{\textit{Mean-field theory.---}}
To understand the two types of phase transitions observed in the nonlinear optical Ising machine, we develop a mean-field (MF) theory where the two-body and four-body spin interaction constants are assumed to be uniform and the fluctuation of the spin magnetization is small. The MF approximation is reliable as the spin interaction in our model is long ranged. Writing $\sigma_i=m + \delta \sigma_i$ with $m\equiv \langle \sigma_i \rangle$ and expanding the spin Hamiltonian \cite{nishimori2001statistical} to linear order in the fluctuations $\delta \sigma_i $, we find 
\begin{equation}
H=(Jm^2N^2+3\gamma Km^4N^4)-h_{eff}\sum_i\sigma_i,
\end{equation}
where $h_{eff}=2JmN+4\gamma Km^3N^3$. The corresponding partition function $Z = \text{Tr}(e^{-H/T})$ is given by 
\begin{equation}
Z = e^{-(Jm^2N^2+3\gamma Km^4N^4)/T} \bigg[2\cosh( h_{eff}/T) \bigg]^N.
\end{equation}
The free energy of the spin system can be obtained with $F\equiv U-TS=-T\ln Z$, where $U$ is the thermal energy and $S$ is the entropy of the system. Substituting the expression for the partition function into the free energy, we find
\begin{eqnarray}
    F &=& -NT\ln2 + Jm^2N^2 + 3\gamma K m^4N^4 \notag \\
    &&- NT\ln\bigg[
    \cosh\left(h_{eff}/{T} \right)
    \bigg].
    \label{eq:F}
\end{eqnarray}
The magnetization $m$ can be calculated after a thermal average
$m=N^{-1}\sum_i\langle \sigma_i\rangle = N^{-1}\sum_i\text{Tr}(\sigma_i e^{-H/T})/Z $
or by taking $dF/dm=0$. A straightforward calculation gives the following self-consistent equation for the magnetization:
\begin{equation}
    m=\tanh[(2JmN+4\gamma Km^3N^3)/T].
\end{equation}
Solving this equation and taking the solution that minimizes the free energy of the spin model yields the ground state magnetization. The MF phase diagram is shown in Fig.~\ref{fig:MFT}(c) which is consistent with that obtained from the Monte-Carlo simulation in the presence of a uniform spin-spin interaction (Fig.~\ref{fig:MFT}(b)). 

We can gain some useful insight into the physics of this system by performing a Taylor expansion of the free energy (Eq.~(\ref{eq:F})) around $m=0$:
\begin{eqnarray}
    F&=&
     - NT\ln2 +N^2J\left(1-\frac{2NJ}{T}\right)m^2 + N^4\bigg[ 
     \frac{4NJ^4}{3T^3} \notag \\
     && - \gamma K\left(
     \frac{8NJ}{T} - 3 
     \right)
     \bigg]m^4 + \mathcal{O}(m^6).
\end{eqnarray}

For $\gamma=0$, the coefficient of the quartic term is always positive. According to Landau's theory of phase transition~\cite{landau2013statistical}, only second-order phase transition is possible unless there are odd terms of magnetization appearing in the free energy expansion (such symmetry-breaking term, for example, can be induced by an external magnetic field). The system exhibits a second-order phase transition when the temperature is larger than a critical value, which can be analytically obtained by letting the coefficient of the quadratic term equal to zero. Hence, we find $T_{c}=2NJ$.

For $\gamma>0$, due to the interplay of the four-body and two-body spin interactions, the quartic term could be positive or negative, giving rise to much richer phase transition phenomena. Specifically, the second-order phase transition persists for smaller positive values of $\gamma$ as the coefficient of the quartic term remains positive at $T\approx T_c$. The phase transition behavior changes dramatically for large enough positive value of $\gamma$ since the coefficient of the quartic term changes sign. By letting it equal to zero at $T=T_c$, we find a critical value of the four-body coefficient $\gamma_c =J/(6N^2K)$. Using the parameters for Fig.~\ref{fig:MFT}(c), we find $\gamma_c\approx 55$ which is very close to the exact numerical value. The discrepancy is due to the ignorance of higher-order terms in the expansion, which is required in order to obtain an accurate estimation of $\gamma_c$. As shown in Fig.~\ref{fig:MFT}(d), the free energy function behaves distinctly at different four-body coefficients. For $\gamma > \gamma_c$, a local minimum at $m=0$ appears in the free energy and the system exhibits a first-order phase transition if this new minimum at $m=0$ becomes a global minimum at a higher temperature. The phase boundary between the ferromagnetic order and paramagnetic order for $\gamma>\gamma_c$ can be estimated using the condition $F(m=0)=F(m=1)$, which gives the simplified form $\gamma = T(\ln 2)/(KN^3)-J/(KN^2)$. Therefore, the critical temperature for the first-order phase transition increases linearly as a function of the four-body coefficient. All these analytical results qualitatively agree with our phase diagram obtained by numerically solving the self-consistent equation, see Fig.~\ref{fig:MFT}(c).

For $\gamma<0$, our system also exhibits a second-order phase transition. However, as shown in Fig.~\ref{fig:MFT}(c), the magnetization in the ferromagnetic phase decreases as $\gamma$ changes from $0$ to $-500$ in order to suppress the four-body spin interaction energy.

\begin{figure}[t!]
\centering
\includegraphics[width=0.48\textwidth]{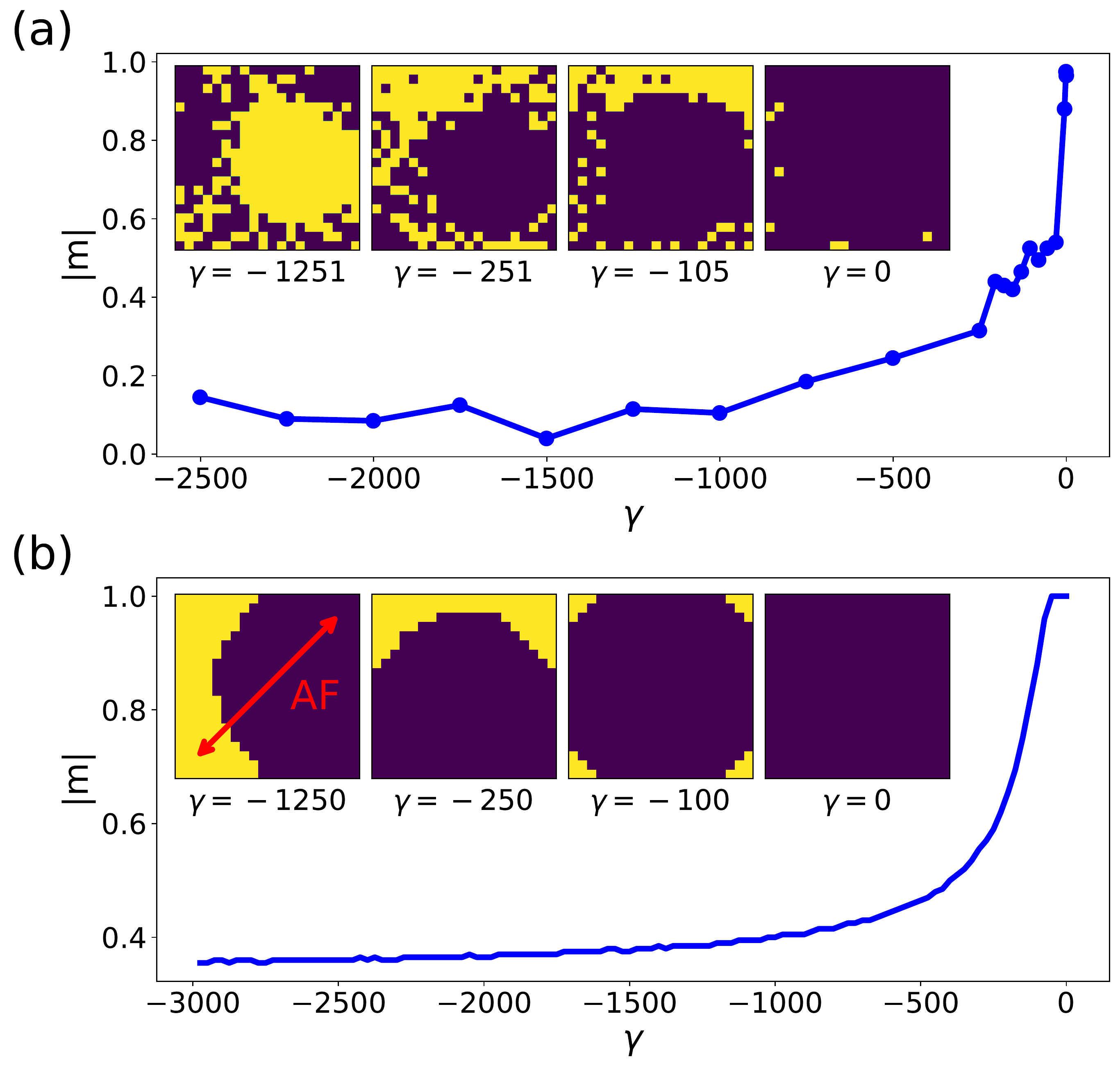}
\caption{Zero temperature magnetic phases of the nonlinear optical Ising machine with a negative four-body coefficient $\gamma<0$. (a) Experimentally measured and (b) numerically obtained magnetization as a function of $\gamma$. The magnetization saturates to a nonzero value for large $|\gamma|$. The inset figures are the spin configurations for some selected $\gamma$ values. The spins located at the opposite corners of the SLM are antiparallel with each other.}
\label{fig:AF}
\end{figure}

\textcolor{blue}{\textit{Nonuniform spin interaction induced spatially structured magnetic phase.---}} For $\gamma<0$, our Monte-Carlo numerical simulation and experimental observation indicate that there is a spatially-structured magnetic phase appearing: the inner part of the SLM exhibits a ferromagnetic phase with all the spins pointing to the same direction and the outer part of the SLM exhibits a long-range anti-ferromagnetic (AF) phase with these spins located at the opposite corners of the SLM pointing to opposite directions (see Fig.~\ref{fig:AF}). This can be understood with the profile of the reduced four-body spin interaction matrix $Q_{ij}$ (Fig.~\ref{fig:ExpSetup}(c)) which depends on the matrix $\xi_{ij}'^f$ whose expression is given below in Eq.~(\ref{eq:K}). The latter matrix has larger elements if $\mathbf{r}_i+\mathbf{r}_j \approx \textbf{0}$. To reduce magnetization, the spins located at the opposite side of the SLM favor opposite directions and the spins next to each other favor the same direction in order to reduce the four-body interaction energy. On the other hand, at the inner part of the SLM where the light intensity is stronger, the two-body interaction dominates and thus a ferromagnetic phase with all the spins point to the same direction is preferred. Consequently, a domain wall that separates the two regions of opposite spins is formed on the SLM (see the inset of Fig.~\ref{fig:AF}(b)).

\textcolor{blue}{\textit{Conclusion.---}} In summary, we have performed systematic investigations of the magnetic phases of the nonlinear optical Ising machine. The exhibited rich phase diagram is a direct consequence of the competition between two-body and four-body spin interactions at different effective temperature. The great flexibility of the nonlinear optical Ising machine may be useful for solving optimization problems where the four-body spin interaction coefficient can be gradually tuned for simulated annealing in order to quickly find the optimal solution \cite{Zoller15,Kanao2021,PRL_Dlaska,susa2020performance}. Our system can be further generalized by considering an anti-ferromagnetic or random two-body spin interactions and including other higher-order spin interactions through sum frequency generation, four-wave mixing, and high harmonic generations. It can be used for the study of q-state Potts model~\cite{blote1979first,honari2020optical} and the development of self-learning Monte-Carlo algorithm~\cite{liu_self-learning_2017}, etc.

\begin{acknowledgments}
This work supported by the ACC-New Jersey under Contract No. W15QKN-18-D-0040.
\end{acknowledgments}

\bibliography{references}

\end{document}